\documentclass[12pt,preprint]{aastex}

\slugcomment{}

\shorttitle{Activity, Metallicity and Radius}
\shortauthors{L\'opez-Morales}

\begin{document}

\title{On the Correlation between the Magnetic Activity Levels, the Metallicities and the Radii of Low-Mass Stars}

\author{Mercedes L\'opez-Morales\altaffilmark{1}}

\email{mercedes@dtm.ciw.edu}

\altaffiltext{1}{Carnegie Fellow. Carnegie Institution of Washington, Department of Terrestrial Magnetism, 5241 Broad Branch Rd. NW, Washington D.C., 20015, USA}

\begin{abstract}

The recent burst in the number of radii measurements of very low-mass stars from eclipsing binaries and interferometry of single stars has opened more questions about what can be causing the discrepancy between the observed radii and the ones predicted by the models. The two main explanations being proposed are a correlation between the radius of the stars and their activity levels or their metallicities. This paper presents a study of such correlations using all the data published to date. The study also investigates correlations between the radii deviation from the models and the masses of the stars. There is no clear correlation between activity level and radii for the single stars in the sample. Those single stars are slow rotators with typical velocities $v_{rot}sini$ $<$ 3.0 km $s^{-1}$. A clear correlation however exists in the case of the faster rotating members of binaries. This result is based on the of X-ray emission levels of the stars. There also appears to be an increase in the deviation of the radii of single stars from the models as a function of metallicity, as previously indicated by Berger et al. (2006). The stars in binaries do not seem to follow the same trend. Finally, the Baraffe et al. (1998) models reproduce well the radius observations below 0.30--0.35 $M_{\sun}$, where the stars become fully convective, although this result is preliminary since almost all the sample stars in that mass range are slow rotators and metallicities have not been measured for most of them. The results in this paper indicate that stellar activity and metallicity play an important role on the determination of the radius of very low-mass stars, at least above 0.35$M_{\sun}$.
\end{abstract}


\section{Introduction} \label{sec:intro}
The two most fundamental parameters of a star are its mass and its radius. For a given stellar mass, models try to reproduce the radius of the star by implementing the best known stellar interior and atmospheric physics. For stars more massive than the Sun, convective interior and radiative atmosphere models reproduce well the observations (e.g. Andersen 1991, 1997). The atmospheres of these stars are dominated by atomic species and their interiors can be closely modeled as an ideal gas. As detailed in the physics of low-mass stars review paper by Chabrier \& Baraffe (2000), below 1$M_{\sun}$ we enter a new physics domain where molecular compounds begin to form in the atmospheres of the stars as their effective temperature drops, and convection expands to the outer layers of the star, until the objects become fully convective below $\sim$ 0.35 $M_{\sun}$. In addition, the interior of the stars becomes denser, and the conditions begin to resemble those of a partially degenerate plasma. Therefore the ideal gas equation of state no longer applies.

Recent measurements of K and M dwarfs radii using both eclipsing binaries and interferometry reveal a disagreement between models and observations. As a general trend, the measured radii appear to be larger than the predictions by the models by factors as large as 20--30\%, in some cases. Two main hypotheses have been suggested to explain this disagreement. The first hypothesis suggests that the discrepancies between models and observations are caused by differences on the level of activity of the stars (L\'opez-Morales \& Ribas 2005; Ribas 2006). The models by Mullan \& McDonald (2001) conclude that the larger radii may be caused by the inhibition of convection in stars with strong magnetic fields, or equivalently, with high rotational velocities.

The second hypothesis suggests that the discrepancies between models and observations are caused by differences in metallicity (Berger et al. 2006). However, the best current models (Baraffe et al. 1998) yield radii values that only differ by about 3\% for metallicities between -0.5--0.0 dex. This suggests that the models could be missing some important source of opacity, as indidated by the authors. A larger range of metallicities in the models is also needed.

The effort of several groups has provided a current sample of 48 radii measurements for K and M main sequence stars\footnote{RW Lac A (Lacy et al. 2005) and HS Aur A (Andersen 1991) have not been included in this study, since they appear to be slightly evolved (see Torres et al. 2006). The M-dwarf in RXJ2130.6+4710 (Maxted et al. 2004) has been also excluded since, being the companion to an evolved white dwarf, it has most likely experienced a common-envelope evolutionary stage.}. Berger et al. (2006), S\'egrensan et al. (2003) and Lane et al. (2001) have measured the radii of 14 single nearby dwarfs through interferometry. The radii of another 34 stars, members of eclipsing binaries, have been measured by different groups (see references in Table 1). Many of the stars in the sample have metallicity estimations and measured X-ray fluxes, a good of magnetic activity indicator (Pevtsov et al. 2003). Therefore, there are now enough data to begin testing possible correlations between the radii of the stars, their activity levels, and their metallicities.

This paper presents a study of the possible correlations between the deviation of the observed stellar radii from the predictions by the models of Baraffe et al. (1998), the X-ray luminosity of the stars, and their estimated metallicities. A study of the radii deviation from the models as a function of mass is also presented. Section 2 describes the data sample used in the analysis. Section 3 shows the analysis of the correlations between the different parameters. Finally, the conclusions of the study are summarized in section 4.

\section{Data Sample} \label{sec:data}
Tables 1 and 2 summarize the main parameters of the current sample of low-mass main sequence stars with radii measurements. The tables include the masses and radii of the stars, their metallicities, their X-ray-to-bolometric luminosity ratios, their rotational velocities, their effective temperatures and their distances. The sample includes 48 objects with masses between 0.092 and 0.960 $M_{\sun}$, covering the entire range of stellar masses between 1 $M_{\sun}$ and the brown dwarf limit. The precision in the mass and radius measurements varies significantly, depending on the apparent brightness of the targets and the observational technique used to derive their parameters. In the case of the components of eclipsing binaries brighter than V = 14, and some secondaries in F--M and G--M binaries studied by Pont et al. (2005) and Bouchy et al. (2005), errors smaller than  2--3\% are reached. For all the other objects, i.e. fainter binaries and single stars with radii measured through interferometry, the error bars are larger, typically 5--10\%. In the binaries, the masses and radii of the stars can be measured directly from the light and radial velocity curves of the systems. In the case of single stars, interferometry provides direct measurements of the stellar radii, but the masses need to be derived using external calibrations. Berger et al. (2006), S\'egransan et al. (2003) and Lane et al. (2001) have derived the masses of the stars in their interferometric studies using the empirical K-band Mass-Luminosity relation from Delfosse et al. (2000) and  the Mass-$M_{K}$ relation by Henry \& McCarthy (1993). The authors adopted errors of 5--10\% in the derived masses to account for photometric and empirical fitting errors. 

Determination of the metallicity of low-mass stars is a difficult task, since as the spectral type of the stars increases, i.e. their effective temperature decreases, absorption bands from molecules such as TiO, VO, $H_{2}$0, CO and CN begin to appear in the spectra, causing the spectral continuum to be underestimated. Efforts to derive accurate metallicities of M-dwarfs are currently underway (see Bonfils et al. 2005, Woolf \& Wallerstein 2005 and Bean et al. 2006). Bonfils et al. (2005) and Woolf \& Wallerstein (2005) have measured metallicities for most of the single stars in the sample. The values that they derive are summarized in column 4 of Table 1. Three of the secondary stars in eclipsing binaries, V818 Tau B, OGLE-TR-34 B, and OGLE-TR-122 B have also reported metallicities, based on the spectral analysis of their more massive primaries (Boesgaard \& Friel 1990; Pont et al. 2005). Those values are also reported in Table 1. In the case of the M-dwarf eclipsing binaries YY Gem, CU Cnc and GU Boo, Torres \& Ribas (2002), Ribas (2003) and L\'opez-Morales \& Ribas (2005) have reported abundances close to solar. The metallicity estimates for CU Cnc and GU Boo are based on the Galactic dynamics of the systems. However those estimations are not strict enough, and have not been considered in the analysis presented in this paper. Torres \& Ribas (2002) have estimated a metallicity of [Fe/H] = 0.1 $\pm$ 0.2 dex for YY Gem, based on metallicity measurements of Castor A and B, of which YY Gem is believed to be a companion. Values of the metallicity of CM Dra have been derived by several authors using both spectral analysis and comparisons with isochrone models (Chabrier \& Baraffe 1995; Leggett et al. 2000; Viti et al. 2002). They obtain values between [Fe/H]= -0.4 and -1.0 dex. For this study I have adopted an average value of [Fe/H] = -0.67 $\pm$ 0.20 dex. Finally, metallicity estimations of V1061 Cyg, FL Lyr, RW Lac, and HS Aur have been derived by Torres et al. (2006). However, those metallicity estimations have not been used, since the are based on model fits and the purpose of this work is to compare models to observations.

Most objects in the sample with apparent magnitudes brighter than V = 14 appear as X-ray sources in the ROSAT All-Sky Bright Star Catalog (Voges et al. 1999), with typical position offsets between the optical sources and their X-ray counterparts of less than 20-30 arcsec. The X-ray counts from the ROSAT Catalog have been converted to fluxes using the equation
\begin{equation}
F_{X} = (5.30 HR + 8.31) 10^{-12} X_{counts}
\end{equation} 

derived by Schmitt et al. (1995), where $F_{X}$ is the X-ray flux of each source, HR is its hardness ratio and $X_{counts}$ is the number of X-ray counts per second. The units of $F_{X}$ are $\it ergs$ $\it s^{-1}$. Those fluxes were then converted to X-ray luminosities using distances from the Hipparcos Catalog (ESA 1997) in the case of the single stars and the binaries FL Lyr and V1061 Cyg. The distances to all the other bright binaries were compiled from Chabrier \& Baraffe (1995), Torres \& Ribas (2002), Ribas (2003), L\'opez-Morales \& Ribas (2005), Lacy et al. (2005), and Torres et al. (2006). All the available distances are listed in column 4 of Table 2. In the case of objects with measured distances, but no X-ray emission detections, X-ray luminosity upper limits have been computed assuming a detection limit of 0.005 counts $s^{-1}$ and HR = -1, the minimum value for the hardness ratio\footnote{This detection limit is based on the source with the lowest number of counts detected by ROSAT (Voges et al. 1999).}. Those objects are BW3 V38, TrES-Her0-07621, 2MASS-J05162881+2607387, 2MASS J04463285+1901432, UNSW-TR-2, V1061 Cyg B, RW Lac and HS Aur. 

Since the stars in the sample cover a wide range of masses, and therefore radii, the X-ray--to--bolometric luminosity ratio, $L_{X}/L_{bol}$, gives a more accurate estimate of the relative activity level differences between the stars. The bolometric luminosities of the stars have been computed from the measured radii and effective temperatures of the objects listed in Tables 1 and 2. 

In the case of the binaries, the X-ray luminosities derived above correspond to the combined contribution from the two stars. Therefore, it was necessary to device a way to estimate the X-ray luminosity of each component. All the binaries in the sample with detected X-ray emission have orbital periods of less than 6 days (in fact, only V818 Tau has P $>$ 3 days), and orbital solutions consistent with e=0.0, i.e. circular orbits (see references in Table 2). The rotational and orbital periods of the stars are therefore expected to be synchronized by tidal interaction. Based on this assumption, the rotational velocity of the stars can be computed from their radii and the orbital period of the systems. This assumption is in fact necessary in the cases where no direct rotational velocity measurements are available (see last paragraph in this section). The secondary stars in the binaries are expected to have slightly slower rotational velocities than the primaries, and one would therefore intuitively expect their magnetic fields to be slightly weaker than the magnetic fields of the primaries. 

X-ray coronal emission has been found to be correlated with stellar rotation in active T-Tauri and late-type main sequence single stars (e.g. Bouvier 1990; Fleming et al. 1989). However, no correlation between those parameters has been found among binaries (Fleming et al. 1989), which are usually faster rotators ($v_{rot}sini$ $\ge$ 10 km$s^{-1}$). This result, attributed to saturation effects, has been corroborated in subsequent studies of $L_{X}/L_{bol}$ emission of G-, K- and M- dwarfs in the Pleiades (Stauffer et al. 1994), the Hyades (Reid et al. 1995; Stauffer et al. 1997), and nearby M-dwarfs (Delfosse et al. 1998). These authors find that in the cluster M-dwarfs the $L_{X}/L_{bol}$ emission saturates at rotational velocities $v_{rot}sini$ $>$ 15 km$s^{-1}$, while for field dwarfs the $L_{X}/L_{bol}$ emission saturates at $v_{rot}sini$ $>$ 5 km$s^{-1}$. In both cases the reported saturation level is $L_{X}/L_{bol}$ $\sim$ 0.003. 

All the components of binaries in the sample with detected X-ray emission have projected rotational velocities larger than $\sim$ 7 km $s^{-1}$, therefore they are expected to fall within the saturated X-ray emission regime, where no correlation between $L_{X}$ and $v_{rot}sini$ has been found. However, to provide a more complete test of any possible correlation between $L_{X}/L_{bol}$ and the radii deviation from the models of low-mass stars, I have estimated the X-ray luminosity of each binary component considering all the observed correlation cases between $L_{X}$ and $v_{rot}sini$. These cases are:

{\it Case 1}: No correlation between $L_{X}$ and $v_{rot}sini$ for all the binaries, as found by (Fleming et al. 1989). In this case it has been assumed that each binary component contributes the same amount of X-ray flux, i.e. $L_{X1}$ = $L_{X2}$ =$L_{X}$/2, where $L_{X}$ is the total X-ray luminosity of the binary measured by ROSAT, and $L_{X1}$ and $L_{X2}$ are, respectively, the luminosities of the primary and the secondary. This is a reasonable approximation for the binaries in this sample, since most of them have components of similar mass.

{\it Case 2}: A linear correlation between $L_{X}$ and $v_{rot}sini$ for all the binaries, i.e. $L_{X}/L_{bol}$ $\propto$ $v_{rot}sini$, as found for single main-sequence stars (Fleming et al. 1989). In this case the X-ray luminosity of each star in the binaries can be computed using the relations
\begin{equation}
L_{X} = L_{X1} + L_{X2}
\end{equation}
and
\begin{equation}
\frac{L_{X1}}{L_{X2}} = \frac{v_{rot1}}{v_{rot2}}
\end{equation}
where $v_{rot1}$ and $v_{rot2}$ are their rotational velocities. The resultant expressions for $L_{X1}$ and $L_{X2}$ are in this case

\begin{equation}
L_{X1} = \frac{v_{r}}{1+v_{r}} L_{X}
\end{equation}
and
\begin{equation}
L_{X2} = \frac{L_{X}}{1+v_{r}}
\end{equation}
where $v_{r} = v_{rot1} / v_{rot2}$. 

{\it Case 3}: A square correlation between $L_{X}$ and $v_{rot}sini$ for all the binaries, i.e. $L_{X}/L_{bol}$ $\propto$ $(v_{rot}sini)^2$, as found for T-Tauri stars (Bouvier 1990). In this case 

\begin{equation}
\frac{L_{X1}}{L_{X2}} = \frac{v_{rot1}^2}{v_{rot2}^2}
\end{equation}

\noindent
and the expressions for $L_{X1}$ and $L_{X2}$ become
\begin{equation}
L_{X1} = \frac{v_{r}^2}{1+v_{r}^2} L_{X}
\end{equation}
and
\begin{equation}
L_{X2} = \frac{L_{X}}{1+v_{r}^2}
\end{equation}

The $L_{X}/L_{bol}$ obtained in each case are summarized in Table 3. The values of $L_{X}/L_{bol}$ listed in Table 1 correspond to {\it Case 1}, since the assumption made in that case is the most consistent with the observational results for binaries.

Finally, Table 2 also includes the estimated rotational velocities for each star. Most of the single stars in the sample have rotational velocity estimates, $v_{rot} sini$, derived directly from the rotational broadening of their spectral lines by Delfosse et al. (1998). In this case, the reported $v_{rot} sini$ values are upper limits, given by the resolution of the spectra used in that work. For the binaries, rotational velocity measurements are only available for some of them, i.e. YY Gem A, B (Torres \& Ribas 2002), GU Boo A, B (L\'opez-Morales \& Ribas 2005), FL LyrB, V1061 Cyg Ab, B, and RW Lac B (Torres et al. 2006). The rotational velocities measurements of these systems are consistent with synchronous rotation in all cases, except RW Lac B. RW Lac has an orbital period of 10.3692 days and a reported eccentricity of e $\sim$ 0.01 (Lacy et al. 2005), therefore, assuming synchronous rotation mat not be correct. The values of $v_{rot} sini$ gien for the binaries in Table 2 correspond to synchronous rotation ($v_{sync} sini$). They have been calculated from the radii of the stars and the orbital period of the binaries published in the literature. In the case of the OGLE binaries reported by Bouchy et al. (2005) and Pont et al. (2005), synchronous rotational velocities have been computed only for the systems with orbital periods $<$ 5 days (see Note $b$ in Table 2).

\section{Mass, X-ray Luminosity, and Metallicity Correlation with Radius}

The Baraffe et al. (1998) models were able to reproduce, for the first time, the Mass-Luminosity and Mass-$M_{V}$ relations of low-mass stars. However, further tests including the Mass-Radius relation and the effect of metallicity and magnetic activity have been inhibited by the lack of observational data.

The new sample of mass, radius, metallicity and stellar activity data compiled in this paper now enables those tests. This section compares the stellar radii predictions of the Baraffe et al. (1998) models to the observed radii of the stars as a function of the other three parameters, i.e. masses, metallicities and stellar activity. The adopted reference model corresponds to an isochrone of age 1 Gigayear, solar metallicity [Fe/H] = 0.0, and mixing length $\alpha$ = 1.0 (standard model)\footnote{The models yield different radii for M $>$ 0.7 $M_{\sun}$ when different values of $\alpha$ are used. 1.0 $R_{\sun}$ for 1.0 $M_{\sun}$ corresponds to $\alpha$ = 1.9 and an age of 4.61 Gyr (Y=0.282). However, the difference in radius with the reference model adopted above is only +0.0013 $R_{\sun}$.}.

Chabrier \& Baraffe (1997) and Baraffe et al. (1997) found that variations of $\alpha$ within a factor of $\sim$ 2 do not significantly alter the result of their models below $\sim$ 0.6 $M_{\sun}$, but those variations become important above that mass. More recently, Yildiz et al. (2006) have found a very definitive correlation between $\alpha$ and stellar mass for masses greater than 0.77 $M_{\sun}$. In particular, they find $\alpha$ = 0.99 $\pm$ 0.03 for V818 Tau B ($\sim$ 0.76$M_{\sun}$) and larger values of $\alpha$ for more massive stars. Based on these results, and the fact that the Baraffe et al. (1998) models only consider values of $\alpha$ = 1.0 and 1.9 (this last one only for masses above 0.7 $M_{\sun}$), the Radius vs. Magnetic Activity and Radius vs. Metallicity correlation tests in \S 3.2 and \S 3.3 only include the stars in the sample with masses $\leq$ 0.77 $M_{\sun}$.

\subsection{Radius versus Mass}

Figure 1 shows the current observational Mass-Radius relation for stars below 1$M_{\sun}$, including all the objects in Table 1. Triangles represent single stars, squares represent low-mass secondaries to eclipsing binaries with primaries $>$ 1$M_{\sun}$, and circles represent the components of eclipsing binaries below 1$M_{\sun}$. The solid line shows the theoretical isochrone model from Baraffe et al. (1998), for an age of 1Gyr, $Z=0.02$, and mixing length $\alpha$ = 1.0. This is the model used as reference throughout this section.

The scatter above 0.30$M_{\sun}$ is significant and the errorbars of most of those measurements certainly need to be improved. However, there are two clear features that can be emphasized at this point. The first one is that the radii measurements for stars below 0.30$M_{\sun}$ seem to agree well with the models. 0.30--0.35$M_{\sun}$ is precisely the mass at which stars are believed to become fully convective. The second feature is that the situation above 0.30$M_{\sun}$ seems completely different. A significant scatter in radius is clear for stars of a given mass. This is most evident between 0.35 and 0.70 $M_{\sun}$, where most measurements clump. The current sample includes 27 stars between 0.35--0.70 $M_{\sun}$, 9 stars above 0.70 $M_{\sun}$, and 12 stars below 0.30 $M_{\sun}$.

\subsection{Radius versus Magnetic Activity}

X-ray emission originating in the magnetically heated corona of the stars is a convenient indicator of stellar magnetic activity. As mentioned before, in single stars magnetic activity has been found to be correlated with stellar rotation (Bouvier 1990; Fleming et al. 1989), while no correlation apparently exists in the case of binaries (Fleming et al. 1989). Fast rotators, such as young stars and tidally spun-up stars in close binaries, generally show higher levels of X-ray emission than slow rotating stars of similar mass. 

As shown in Table 2, there are two different populations of rotators in the data sample. All the single stars are slow rotators, with $v_{rot} sini <$ 2.9 $km s^{-1}$. The components of binaries, on the other hand, rotate with velocities ranging from $v_{rot}$ $\sim$ 2.0 to 130 $km s^{-1}$. Those differences in rotational velocities translate into differences in magnetic activity levels of the order of a hundred (see column 5 in Table 1).

The top diagram in Figure 2 shows the fractional deviation of the radii of the stars from the 1Gyr, Z = 0.02 Baraffe et al. (1998) model as a function of $L_{X}/L_{bol}$ for the single stars in the sample. The bottom diagram in that figure and the two diagrams in Figure 3 show the same fractional radii deviation for the stars in binaries. Fig 2--bottom represents {\it Case 1} in \S 2. The two diagrams in Figure 3 represent, respectively, {\it Case 2} (top) and {\it Case 3} (bottom). Notice that the values of the x-axis in the binary diagrams are approximately a hundred times larger than the values in Fig 2--top. This reflects the enhanced magnetic astivity levels in the components of binaries. The value of $L_{X}/L_{bol}$ measured for GJ551 in Fig 2--top is 2.728 $10^{-4}$ $\pm$ 6.508 $10^{-5}$, that is, a factor of 20-30 times higher than all the other single stars in the sample. That value has been rescaled by a factor of ten to include that point in the figure.

The main conclusions that one can derive from the diagrams in Figures 2 and 3 are that there is no apparent correlation between the level of magnetic activity and the deviation of the radius from the models in the case of single stars. However, a clear correlation exists in the case of the more magnetically active components of binaries. These conclusions are endorsed by an statistical analysis of the data samples by computing the significance of the correlations in each figure using Pearson's correlation coefficient $r$. The resultant values of $r$ are $r$ = -0.115, for the data in Fig 2--top, and $r$ = 0.890, 0.899, and 0.901, for the data in Fig 2--bottom, Fig 3--top, and Fig 3--bottom, respectively. These results indicate that the null hypothesis, which states that $L_{X}/L_{bol}$ and the radii deviation from the models are uncorrelated, can be only rejected at the $<$ 30$\%$ confidence level for single stars, while for the stars in binaries the null hypothesis can be rejected at the $>$ 99$\%$ confidence level in all the cases. The dotted lines in Figure 2--bottom and Figure 3 show the best least square fits to the data, using only the binaries with detected X-ray emission, and taking into account the errorbars in both quantities. The results of those least square fits are in each case

\begin{description}
\item[]  Case 1: ($R_{obs}$ - $R_{mod}$)/$R_{mod}$ $\propto$ 74.709 $L_{X}/L_{bol}$ ; $\chi^{2}$ = 8.41
\item[]  Case 2: ($R_{obs}$ - $R_{mod}$)/$R_{mod}$ $\propto$ 73.430 $L_{X}/L_{bol}$ ; $\chi^{2}$ = 8.74   
\item[]  Case 3: ($R_{obs}$ - $R_{mod}$)/$R_{mod}$ $\propto$ 71.690 $L_{X}/L_{bol}$ ; $\chi^{2}$ = 9.11
\end{description}

The data used in the least square fits only includes stars with M $\leq$ 0.77$M_{\sun}$ to avoid biases introduced by a changing mixing-length parameter (see \S 3). When the data for stars more massive than 0.77$M_{\sun}$ (FL Lyr B and V1061 Cyg Ab) are included, the slopes of the correlations above change to 65.157 (case 1), 62.511 (case 2) and 59.998 (case 3). The Pearson's correlation coefficients become $r$ = 0.833, 0.832, 0.829, and the null hypothesis can be still rejected at the same confidence levels obtained above. The dot-dashed lines in Figures 2 and 3 show the best least square fits including stars $>$ 0.77 $M_{\sun}$.

\subsection{Radius versus Metallicity}

Although there is no clear correlation between the magnetic activity levels and radii deviation from the models in the case of single stars (Figure 2--${\it top}$) the dispersion in radius is quite large. That dispersion could be attributed to metallicity differences, as shown below.

The open circles in Fig. 2--top show stars with [Fe/H] $<$ -0.25, while the filled circles show stars with metallicities higher than that value. The crosses correspond to stars with no available metallicity measurements. With the exception of GJ380, all stars with [Fe/H] $>$ -0.25 show larger radii deviations from the models than stars with [Fe/H] $<$ -0.25. There are also signs of a gradient between the metallicity of the stars and their radii. That gradient is more clearly illustrated in Figure 4.

Figure 4 represents the relative deviation of the radii of the stars from the models as a function of metallicity. The {\it top} diagram corresponds to single stars. The {\it bottom} diagram shows the metallicity estimations for the low-mass secondaries in binaries with primaries $>$ 1$M_{\sun}$, V818 Tau B, OGLE-TR-34 B, and OGLE-TR-122 B, and the M-dwarf binaries YY Gem and CM Dra. The radii deviation from the models increases with metallicity in the case of single stars, as previously noticed by Berger et al. (2006). However the stars in binaries do not appear to follow that same trend. As with the Radius vs. Magnetic Activity correlation in \S 3.2, the conclusions in this section have been statistically checked by computing the significant of the correlations using Pearson's correlation coefficient. The values of $r$ are in this case $r$ = 0.504 for single stars, and $r$ = -0.209 for the binaries. The null hypothesis assuming that [Fe/H] and the relative radius deviation from the models are uncorrelated can be rejected at the 90$\%$ confidence level in the case of single stars. In the case of the binaries the null hypothesis cannot be discarded. The dotted lines in Figure 4 show the best least square fit for each sample, taking into account the errorbars in both quantities. The results of those least square fits are

\begin{description}
\setlength{\parskip}{.05ex}
\item[]  For Single Stars: ($R_{obs}$ - $R_{mod}$)/$R_{mod}$ $\propto$ 0.20 [Fe/H] ; $\chi^{2}$ = 10.70
\item[]  For Binaries: ($R_{obs}$ - $R_{mod}$)/$R_{mod}$ $\propto$ 0.04 [Fe/H] ; $\chi^{2}$ = 10.18   
\end{description}

\noindent
In this case, all the stars in both samples have masses below 0.77$M_{\sun}$.

\section{Conclusions} \label{sec:concl}

This paper presents a compilation of all the current mass-radius measurements of main sequence stars below 1 $M_{\sun}$, in addition to all the existing information about their X-ray emission levels and metallicities. The goal has been to find any correlation between the radii of the stars, their masses, their magnetic activity levels, and their metallicities. The stellar radii have been compared to the Baraffe et al. (1998) isochrone model for an age of 1 Gigayear, solar metallicity (Z=0.02), and mixing length $\alpha$ = 1.0.

The comparison of the observed Mass-Radius relation to the one predicted by the Baraffe et al. (1998) model shows good agreement for stars below 0.30--0.35$M_{\sun}$. Above those masses, the deviation of the measured radii from the values predicted by the models is significant. In some cases that difference is as large as 30$\%$. The result for masses below 0.35 $M_{\sun}$ can be a bit misleading, since all the objects in that mass range have rotational  velocities $v_{rot} sini$ $<$ 10 km$s^{-1}$ (except OGLE-TR-5B, with $v_{rot} sini$ = 16.5 km$s^{-1}$). More radii measurements of active stars in this mass regime are necessary before we can arrive to any conclusion about a radius--magnetic activity correlation for those stars.

Above 0.35$M_{\sun}$, the sample includes stars with a wide range of magnetic activity levels and metallicities. There is in this case a significant scatter in radius for any given stellar mass. That scatter can be attributed to differences in the magnetic activity level and the metallicity of the stars, as shown in \S 3. The radii of the stars do not appear to be correlated with magnetic activity in the case of slow rotators. Slow rotators are usually single stars, old enough to have had time to slow down rotationally. The typical rotational velocities of these objects are less than 5 km $s^{-1}$. However, there is a clear correlation between radius and magnetic activity in the case of faster rotating, more magnetically active, stars. It has been postulated in previous papers (e.g. L\'opez-Morales \& Ribas 2005) that the significant areal coverage of active regions (spots) on the surface of magnetically active stars changes the overall photospheric temperature of the stars; an effect that they could compensate by increasing their radii to conserve their total radiative fluxes. Torres et al. (2006) have compared the radii of two fast rotators (V1061 Cyg Ab and FL Lyr B) and two slow rotators (RW Lac A and HS Aur A) of similar mass and find the fast rotators to have larger radii than what the models predict. The theoretical work by Mullan \& McDonald  (2001) concludes that inhibition of convection in fast rotating stars may be the cause of their larger radii. This might be the case not only for low-mass stars in close binaries, but also for young single stars and T Tauri objects, which are usually very active. 

Metallicity also appears to be playing a role in the radii of the stars. There is a clear correlation between the size of the stars and how metal rich they are. This correlation, already noted by Berger et al. (2006), appears to apply to single stars, but not to the components of binaries. However, this last conclusion is still premature, since only a handful of the low-mass stars in binaries have metallicity estimations. Furthermore, most of those metallicity estimations are indirect. Direct metallicity measurements of the components of low-mass binaries are needed to determine any correlation with the radii of the stars. An increasing abundance of metals would have the effect of enhancing the number density of molecular compounds in the atmospheres of the stars, making it harder for the radiation to escape. The response of the stars might be similar to the case of high magnetic activity, i.e. they could increase their radii to conserve their radiative fluxes.

\acknowledgments

The author acknowledges research support from the Carnegie Institution of Washington through a Carnegie Fellowship. The author is also thankful to referee J. Southworth for a number of useful comments and suggestions. The material in this paper has been compiled using  the SIMBAD database, operated at CDS, Strasbourg, France and the NASA Astrophysics database system. This research has been partially supported by the National Science Foundation through grant AST 00-94289.

\clearpage
\pagestyle{empty}
\setlength{\voffset}{-25mm}

\begin{table}[t]
\scriptsize
\centering
\caption{Mass, radius, metallicity and X-ray-to-Bolometric luminosity ratio of the stars in the sample}
\label{tab:params_01} 
\begin{tabular}{lccccl}
\hline\hline
Star&M &R &[Fe/H] &$L_{x}/L_{bol}$& Source\tablenotemark{1}\\
&($M_{\sun}$)&($R_{\sun}$)&(dex)&&\\
\hline
GJ 15 A ......................................&0.4040$\pm$0.0404&0.379$\pm$0.006&-0.46$\pm$0.20&2.029E-05$\pm$3.627E-06&1,15\\
GJ 514 ........................................&0.5260$\pm$0.0526&0.611$\pm$0.043&-0.24$\pm$0.20&1.509E-05$\pm$6.070E-06&1,15\\
GJ 526 ........................................&0.5020$\pm$0.0502&0.493$\pm$0.033&-0.31$\pm$0.20&2.653E-06$\pm$4.583E-07&1,15\\
GJ 687 ........................................&0.4010$\pm$0.0401&0.492$\pm$0.038&0.11$\pm$0.20&9.433E-06$\pm$2.027E-06&1,15\\
GJ 752 A ....................................&0.4840$\pm$0.0484&0.526$\pm$0.032&-0.05$\pm$0.20&4.088E-06$\pm$7.631E-07&1,15\\
GJ 880 ........................................&0.5860$\pm$0.0586&0.689$\pm$0.044&-0.04$\pm$0.20&6.296E-06$\pm$2.627E-06&1,15\\
GJ 205 ........................................&0.631$\pm$0.031&0.702$\pm$0.063&0.21$\pm$0.13&1.500E-05$\pm$4.533E-06&2,15\\
GJ 887 ........................................&0.503$\pm$0.025&0.491$\pm$0.014&-0.22$\pm$0.09&5.245E-06$\pm$1.762E-06&2,15\\
GJ 191 ........................................&0.281$\pm$0.014&0.291$\pm$0.025&-0.90$\pm$0.20&9.904E-06$\pm$2.427E-06&2,15\\
GJ 551 ........................................&0.123$\pm$0.006&0.145$\pm$0.011&\nodata&2.728E-04$\pm$6.508E-05&2\\
GJ 699 ........................................&0.158$\pm$0.008&0.196$\pm$0.008&-0.50$\pm$0.30&2.880E-06$\pm$3.336E-07&3,17\\
GJ 411 ........................................&0.403$\pm$0.020&0.393$\pm$0.008&-0.42$\pm$0.07&7.169E-06$\pm$1.384E-06&3,15\\
GJ 380 ........................................&0.670$\pm$0.033&0.605$\pm$0.020&-0.03$\pm$0.14&8.954E-06$\pm$2.063E-06&3,15\\
GJ 105 A ....................................&0.790$\pm$0.039&0.708$\pm$0.050&\nodata&2.082E-06$\pm$8.738E-07&3\\
\hline
CM Dra A ...................................&0.2307$\pm$0.0010&0.2516$\pm$0.0020&-0.67$\pm$0.20&6.518E-04$\pm$1.037E-04&4,16\\
CM Dra B ...................................&0.2136$\pm$0.0010&0.2347$\pm$0.0019&-0.67$\pm$0.20&7.858E-04$\pm$1.259E-04&4,16\\
YY Gem A ..................................&0.5992$\pm$0.0047&0.6191$\pm$0.0057&0.10$\pm$0.20&1.326E-03$\pm$1.460E-04&5\\
YY Gem B ..................................&0.5992$\pm$0.0047&0.6191$\pm$0.0057&0.10$\pm$0.20&1.326E-03$\pm$1.460E-04&5\\
CU Cnc A ...................................&0.4333$\pm$0.0017&0.4317$\pm$0.0052&\nodata&8.457E-04$\pm$1.755E-04&6\\
CU Cnc B ...................................&0.3980$\pm$0.0014&0.3908$\pm$0.0094&\nodata&1.079E-03$\pm$2.305E-04&6\\
GU Boo A ...................................&0.610$\pm$0.007&0.623$\pm$0.016&\nodata&1.191E-03$\pm$4.991E-04&7\\
GU Boo B ...................................&0.599$\pm$0.006&0.620$\pm$0.020&\nodata&1.347E-03$\pm$5.688E-04&7\\
BW3 V38 A .................................&0.44$\pm$0.07&0.51$\pm$0.04&\nodata&$<$ 1.08E-03&8\\
BW3 V38 B .................................&0.41$\pm$0.09&0.44$\pm$0.06&\nodata&$<$ 1.53E-03&8\\
TrES-Her0-07621 A ......................&0.493$\pm$0.003&0.453$\pm$0.060&\nodata&$<$ 1.90E-04&9\\
TrES-Her0-07621 B ......................&0.489$\pm$0.003&0.452$\pm$0.050&\nodata&$<$ 1.35E-04&9\\
2MASS J05162881+2607387 A ....&0.787$\pm$0.012&0.788$\pm$0.015&\nodata&$<$ 7.70E-04&10\\
2MASS J05162881+2607387 B ....&0.770$\pm$0.009&0.817$\pm$0.010&\nodata&$<$ 7.48E-04&10\\
2MASS J04463285+1901432 A .................................&0.47$\pm$0.05&0.57$\pm$0.02&\nodata&$<$ 1.95E-03&11\\
2MASS J04463285+1901432 B .................................&0.19$\pm$0.02&0.21$\pm$0.01&\nodata&$<$ 2.44E-02&11\\
UNSW-TR-2 A  ............................&0.529$\pm$0.035&0.641$\pm$0.05&\nodata&\nodata&12\\
UNSW-TR-2 B ............................&0.512$\pm$0.035&0.608$\pm$0.06&\nodata&\nodata&12\\
\hline
V818 Tau B .................................&0.7605$\pm$0.0062&0.768$\pm$0.010&0.13$\pm$0.02&4.025E-04$\pm$7.738E-05&5,18\\
FL Lyr B .....................................&0.960$\pm$0.012&0.962$\pm$0.028&\nodata&2.819E-04$\pm$5.975E-05&19,20\\
V1061 Cyg Ab .............................&0.9315$\pm$0.0068&0.974$\pm$0.020&\nodata&2.373E-04$\pm$7.190E-05&19\\
V1061 Cyg B .............................&0.925$\pm$0.036&0.870$\pm$0.087&\nodata&$<$ 1.72E-05&19\\
RW Lac B .............................&0.870$\pm$0.004&0.964$\pm$0.004&\nodata&$<$ 1.07E-05&19,21\\
HS Aur B .............................&0.879$\pm$0.017&0.873$\pm$0.024&\nodata&$<$ 4.66E-06&19,22\\
OGLE-TR-5 B .............................&0.271$\pm$0.035&0.263$\pm$0.012&\nodata&\nodata&14\\
OGLE-TR-6 B .............................&0.359$\pm$0.025&0.393$\pm$0.018&\nodata&\nodata&14\\
OGLE-TR-7 B .............................&0.281$\pm$0.029&0.282$\pm$0.013&\nodata&\nodata&14\\
OGLE-TR-18 B ...........................&0.387$\pm$0.049&0.39$\pm$0.04&\nodata&\nodata&14\\
OGLE-TR-34 B ...........................&0.509$\pm$0.038&0.435$\pm$0.033&0.32$\pm$0.31&\nodata&14,15\\
OGLE-TR-78 B ...........................&0.243$\pm$0.015&0.24$\pm$0.013&\nodata&\nodata&13\\
OGLE-TR-106 B .........................&0.116$\pm$0.021&0.181$\pm$0.013&\nodata&\nodata&13\\
OGLE-TR-120 B .........................&0.47$\pm$0.04&0.42$\pm$0.02&\nodata&\nodata&13\\
OGLE-TR-122 B .........................&0.092$\pm$0.009&0.120$\pm$0.019&0.15$\pm$0.36&\nodata&13\\
OGLE-TR-125 B .........................&0.209$\pm$0.033&0.211$\pm$0.027&\nodata&\nodata&13\\
\hline\hline
\end{tabular}
\tablenotetext{1}{[1] Berger et al. (2006), [2] S\'egransan et al. (2003), [3] Lane et al. (2001), [4] Metcalfe et al. (1996), [5] Torres \& Ribas (2002), [6] Ribas (2003), [7] L\'opez-Morales \& Ribas (2005), [8] Maceroni \& Montalb\'an (2004), [9] Creevey et al. (2005), [10] Bayless \& Orosz (2006), [11] Hebb et al. (2006), [12] Young et al. (2006), [13] Pont et al. (2005), [14] Bouchy et al. (2005), [15] Bonfils et al. (2005), [16] average of values derived by Chabrier \& Baraffe (1995), Legget et al. (2000) and Viti et al. (2002), [17] Jones et al. (2002), [18] Boesgaard \& Friel (1990), [19] Torres et al. (2006), [20] Popper et al. (1986), [21] Lacy et al. (2005), [22] Andersen (1991).}
\end{table}

\begin{table}[t]
\scriptsize
\centering
\caption{Rotational velocity, effective temperature and distance of the stars in the sample}
\label{tab:params_02} 
\begin{tabular}{lrccl}
\hline\hline
Star&$v_{rot}sini$&$T_{eff}$&Distance& Source\tablenotemark{a}\\
&($km s^{-1}$)&(K)&(parsecs)&\\
\hline
GJ 15 A ......................................&$<$ 2.9&3747 $\pm$ 100&3.57 $\pm$ 1.4E-05&1,2,3\\
GJ 514 ........................................&$<$ 2.9&3377 $\pm$ 100&7.63 $\pm$ 9.7E-05&1,2,3\\
GJ 526 ........................................&$<$ 2.9&3662 $\pm$ 100&5.43 $\pm$ 4.7E-05&1,2,3\\
GJ 687 ........................................&$<$ 2.8&3142 $\pm$ 100&4.53 $\pm$ 1.7E-05&1,2,3\\
GJ 752 A ....................................&$<$ 2.6&3390 $\pm$ 120&5.87 $\pm$ 6.5E-05&1,2,3\\
GJ 880 ........................................&$<$ 2.8&3373 $\pm$ 100&6.89 $\pm$ 7.1E-05&1,2,3\\
GJ 205 ........................................&$<$ 2.9&3520 $\pm$ 170&5.69 $\pm$ 4.7E-05&1,4,3\\
GJ 887 ........................................&\nodata&3626 $\pm$ 56&3.29 $\pm$ 8.2E-06&4,3\\
GJ 191 ........................................&\nodata&3570 $\pm$ 156&3.92 $\pm$ 1.13E-05&4,3\\
GJ 551 ........................................&\nodata&3042 $\pm$ 117&1.29 $\pm$ 9.8E-06&4,3\\
GJ 699 ........................................&$<$ 2.8&3163 $\pm$ 65&1.82 $\pm$ 8.3E-06&1,4,3\\
GJ 411 ........................................&$<$ 2.9&3570 $\pm$ 42&2.55 $\pm$ 5.4E-06&1,4,3\\
GJ 380 ........................................&2.8&3950 $\pm$ 161&4.87 $\pm$ 1.6E-05&1,21,3\\
GJ 105 A ....................................&\nodata&4714 $\pm$ 67&7.21 $\pm$ 5.6E-05&21,3\\
\hline
CM Dra A ...................................&10.03 $\pm$ 0.08&3360 $\pm$ 100&15.9 $\pm$ 0.1&5,6\\
CM Dra B ...................................&9.35 $\pm$ 0.08 &3320 $\pm$ 100&15.9 $\pm$ 0.1&5,6\\
YY Gem A ..................................&38.40 $\pm$ 0.40&3820 $\pm$ 100&14.90 $\pm$ 1.0E-04&7\\
YY Gem B ..................................&38.40 $\pm$ 0.40&3820 $\pm$ 100&14.90 $\pm$ 1.0e-04&7\\
CU Cnc A ...................................&7.86 $\pm$ 0.09&3160 $\pm$ 150&12.81 $\pm$ 5.0E-03&8\\
CU Cnc B ...................................&7.11 $\pm$ 0.17&3125 $\pm$ 150&12.81 $\pm$ 5.0E-03&8\\
GU Boo A ...................................&64.37$\pm$ 1.65&3920 $\pm$ 130&140 $\pm$ 8&9\\
GU Boo B ...................................&64.06$\pm$ 2.07&3810 $\pm$ 130&140 $\pm$ 8&9\\
BW3 V38 A .................................&129.5 $\pm$ 10.2&3500&$\sim$ 400&10\\
BW3 V38 B .................................&111.7 $\pm$ 15.3&3448&$\sim$ 400&10\\
TrES-Her0-07621 A ......................&20.28 $\pm$ 2.67&3500&118 $\pm$ 13&11\\
TrES-Her0-07621 B ......................&20.24 $\pm$ 2.24&3395&118 $\pm$ 13&11\\
2MASS J05162881+2607387 A ....&15.31 $\pm$ 0.29&4200 $\pm$ 200&753 $\pm$ 34&12\\
2MASS J05162881+2607387 B ....&15.87 $\pm$ 0.19&4154 $\pm$ 200&753 $\pm$ 34&12\\
2MASS J04463285+1901432 A .................................&46.02 $\pm$ 1.61&3320 $\pm$ 150&$\sim$ 542&13\\
2MASS J04463285+1901432 B .................................&16.95 $\pm$ 0.81&2910 $\pm$ 150&$\sim$ 542&13\\
UNSW-TR-2 A  ............................&15.21 $\pm$ 1.19&\nodata&169 $\pm$ 14&14\\
UNSW-TR-2 B ............................&14.42 $\pm$ 1.42&\nodata&169 $\pm$ 14&14\\
\hline
V818 Tau B .................................&6.90 $\pm$ 0.09&4220 $\pm$ 150&46.8 $\pm$ 0.4&7,3\\
FL Lyr B .....................................&22.28 $\pm$ 0.65&5300 $\pm$ 100&130.04 $\pm$ 0.025&15,16\\
V1061 Cyg Ab .............................&20.97 $\pm$ 0.43&5300 $\pm$ 150&160.00 $\pm$ 0.03  &15\\
V1061 Cyg B .............................&\nodata&5670$ \pm$ 150&160.00 $\pm$ 0.03&15\\
RW Lac B\tablenotemark{b} .............................&\nodata&5560 $\pm$ 150&190 $\pm$ 10&15,19\\
HS Aur B .............................&\nodata&5200 $\pm$ 75&99.5 $\pm$ 0.05&15,20,3\\
OGLE-TR-5 B .............................&16.39 $\pm$ 0.75&\nodata&\nodata&17\\
OGLE-TR-6 B .............................&4.37 $\pm$ 0.20&\nodata&\nodata&17\\
OGLE-TR-7 B .............................&5.24 $\pm$ 0.24&\nodata&\nodata&17\\
OGLE-TR-18 B ...........................&8.77 $\pm$ 0.90&\nodata&\nodata&17\\
OGLE-TR-34 B\tablenotemark{b} ...........................&\nodata&\nodata&\nodata&\\
OGLE-TR-78 B\tablenotemark{b} ...........................&\nodata&\nodata&\nodata&\\
OGLE-TR-106 B .........................&3.59 $\pm$ 0.26&\nodata&\nodata&18\\
OGLE-TR-120 B\tablenotemark{b} .........................&\nodata&\nodata&\nodata&\\
OGLE-TR-122 B\tablenotemark{b} .........................&\nodata&\nodata&\nodata&\\
OGLE-TR-125 B .........................&2.00 $\pm$ 0.26&\nodata&\nodata&18\\
\hline\hline
\end{tabular}
\tablenotetext{a}{[1] Delfosse et al. (1998), [2] Berger et al. (2006), [3] Hipparcos Catalog (ESA 1997), [4] S\'egransan et al. (2003), [5] Metcalfe et al. (1996), [6] Chabrier \& Baraffe (1995), [7] Torres \& Ribas (2002), [8] Ribas (2003), [9] L\'opez-Morales \& Ribas (2005), [10] Maceroni \& Montalb\'an, [11] Creevey et al. (2005), [12] Bayless \& Orosz (2006), [13] Hebb et al. (2006), [14] Young et al. (2006), [15] Torres et al. (2006), [16] Popper et al. (1986), [17] Bouchy et al. (2005), [18] Pont et al. (2005), [19] Lacy et al. (2005), [20] Andersen (1991), [21] Ram\'irez \& Mel\'endez (2004)}
\tablenotetext{b}{Low-mass secondaries in binaries with orbital periods $>$ 5 days. Three of these binaries show eccentricities e $>$ 0.0 (Pont et al. 2005). Therefore, the rotational velocities of the stars can not be computed from the assumption that the systems are tidally locked. This is consistent with the theoretical result from Zahn (1989), that predicts larger orbital circularization timescales for binaries with small mass ratios. Pont et al. (2005) conclude from their data that the typical circularization period for these systems is 5 days. The binary OGLE-TR-125A-B has a period of 5.3039 days, but eccentricity e = 0.0 $\pm$ 0.01. therefore its orbit is circularized.}
\end{table}

\begin{table}[t]
\scriptsize
\centering
\caption{X-ray-to-bolometric luminosity ratio of the stars in each binary with detected X-ray emission. {\it Case 1} corresponds to no correlation between $L_{X}$ and $v_{rot}$sini, {\it Case 2} corresponds to $L_{X}$ $\propto$ $v_{rot}$sini, and {\it Case 3} corresponds to $L_{X}$ $\propto$ ($v_{rot}$sini)$^2$.}
\label{tab:params_03} 
\begin{tabular}{lccc}
\hline\hline
&Case 1&Case 2&Case 3\\
Star&$L_{X}/L_{bol}$&$L_{X}/L_{bol}$&$L_{X}/L_{bol}$\\
\hline
CM Dra A ...................................&6.518E-04 $\pm$ 1.037E-04& 6.745E-04 $\pm$	1.073E-04 & 6.971E-04 $\pm$ 1.111E-04 \\
CM Dra B ...................................&7.858E-04 $\pm$ 1.259E-04& 7.585E-04 $\pm$	1.216E-04 & 7.313E-04 $\pm$ 1.175E-04 \\
YY Gem A ..................................&1.326E-03 $\pm$ 1.460E-04& 1.326E-03 $\pm$	1.462E-04 & 1.326E-03 $\pm$ 1.470E-04 \\
YY Gem B ..................................&1.326E-03 $\pm$ 1.460E-04& 1.326E-03 $\pm$	1.462E-04 & 1.326E-03 $\pm$ 1.470E-04 \\
CU Cnc A ...................................&8.457E-04 $\pm$ 1.755E-04& 8.877E-04 $\pm$	1.846E-04 & 9.295E-04 $\pm$ 1.943E-04 \\
CU Cnc B ...................................&1.079E-03 $\pm$ 2.305E-04& 1.025E-03 $\pm$	2.195E-04 & 9.719E-04 $\pm$ 2.096E-04  \\
GU Boo A ...................................&1.191E-03 $\pm$ 4.991E-04& 1.194E-03 $\pm$	5.009E-04 & 1.197E-03 $\pm$ 5.039E-04 \\
GU Boo B ...................................&1.347E-03 $\pm$ 5.688E-04& 1.344E-03 $\pm$	5.681E-04 & 1.341E-03 $\pm$ 5.688E-04 \\
V818 Tau B .................................&4.025E-04 $\pm$ 7.738E-05& 3.706E-04 $\pm$	7.140E-05 & 3.392E-04 $\pm$ 6.578E-05 \\
FL Lyr B .....................................&2.819E-04 $\pm$ 5.975E-05& 2.417E-04 $\pm$ 5.148E-05 & 2.031E-04 $\pm$ 4.408E-05 \\
V1061 Cyg Ab .............................&2.373E-04 $\pm$ 7.190E-05& 1.786E-04 $\pm$ 5.416E-05 & 1.266E-04 $\pm$ 3.859E-05 \\
\hline\hline
\end{tabular}
\end{table}

\clearpage
\pagestyle{plaintop}
\setlength{\voffset}{0mm}

\begin{figure}[t]
\epsscale{1.0}
\plotone{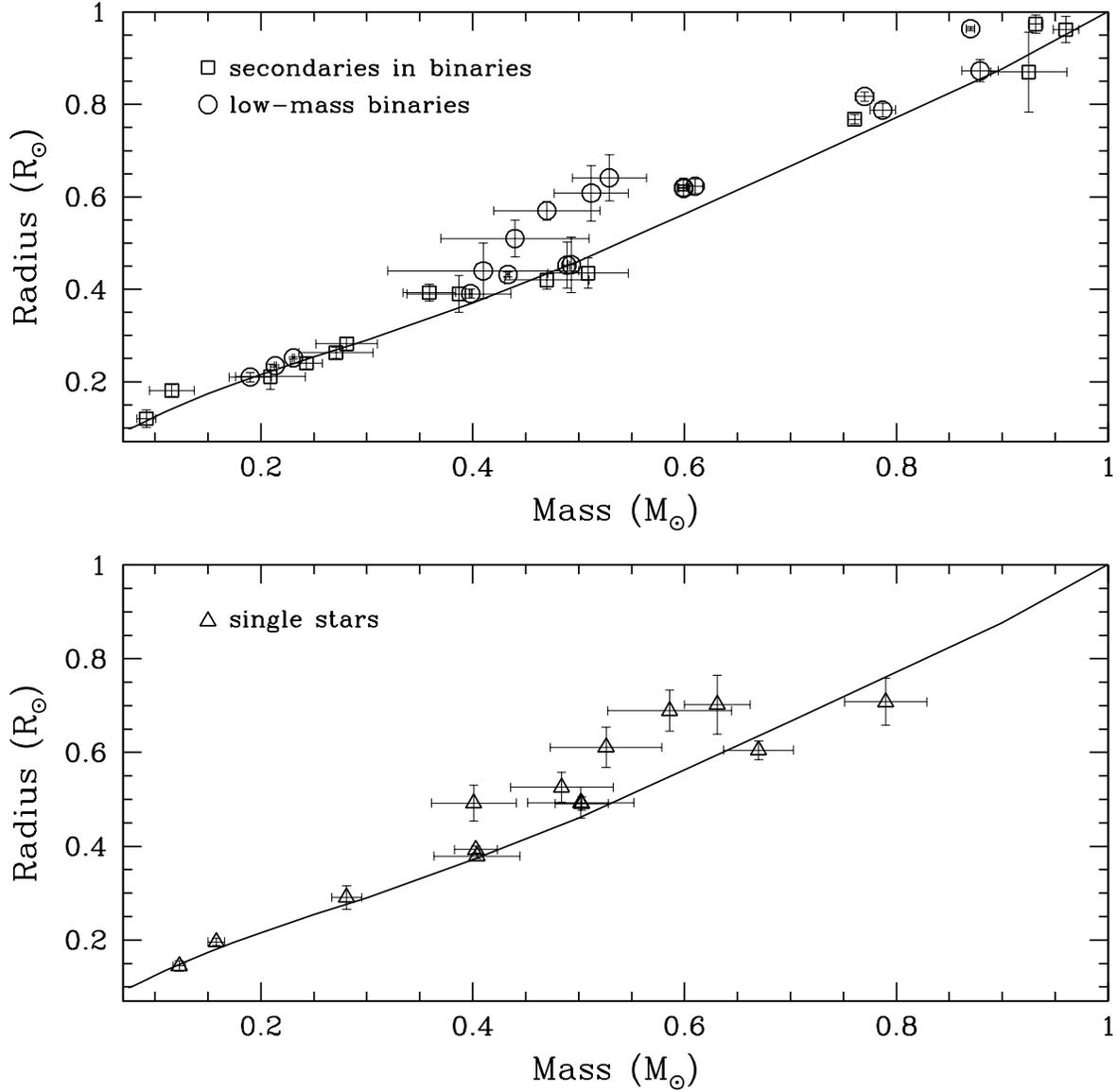}
\caption{Current observational Mass-Radius relation for stars below 1$M_{\sun}$ (objects in Table 1). The {\it top} figure shows all the data from low-mass secondaries to eclipsing binaries with primaries $>$ 1$M_{\sun}$ (squares), and the components of eclipsing binaries below 1$M_{\sun}$ (circles). The {\it bottom} figure shows all the measurements from single stars (triangles). The solid line in both figures represents the theoretical isochrone model from Baraffe et al. (1998), for an age of 1Gyr, $Z=0.02$, and mixing length $\alpha$ = 1.0 (standard model).}
\label{fig:lcs}
\end{figure}

\begin{figure}[t]
\epsscale{1.0}
\plotone{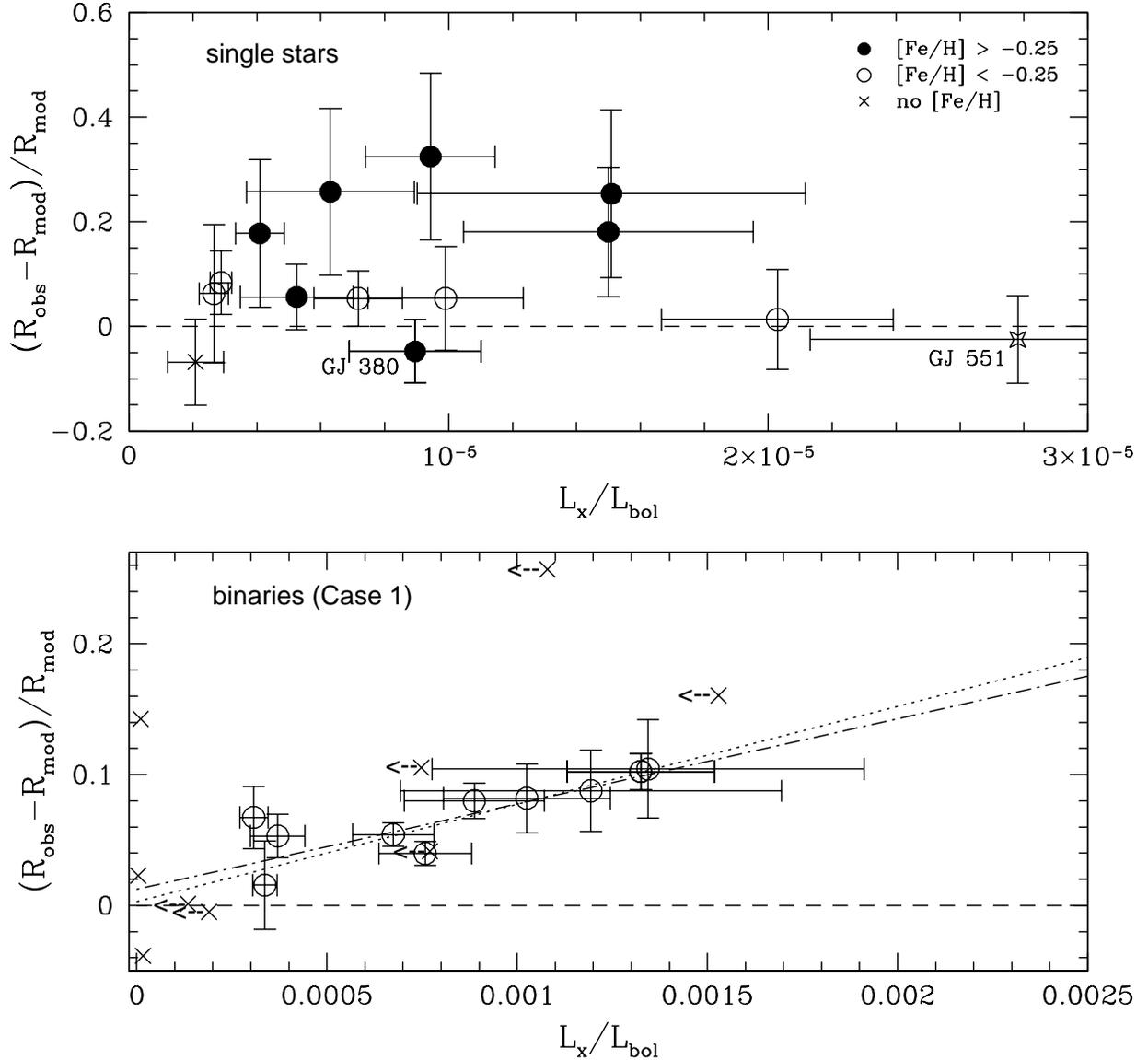}
\caption{${\it Top}$ -- Fractional deviation of the radius of the single stars in the sample from the 1Gyr, Z = 0.02, $\alpha$ = 1.0 Baraffe et al. (1998) model as a function of $L_{X}/L_{bol}$. Filled circles show stars with [Fe/H] $>$ -0.25 dex and open circles show stars with [Fe/H] $<$ -0.25 dex. Stars with no available metallicity estimations are shown as crosses. ${\it Bottom}$ -- Fractional deviation of the radius of the stars in binaries from the 1Gyr, Z = 0.02, $\alpha$ = 1.0 Baraffe et al. (1998) model as a function of $L_{X}/L_{bol}$ for {\it Case 1}, i.e. assuming no correlation between $L_{X}/L_{bol}$ and $v_{rot}$sini. The open circles show the component of binaries with X-ray emission detected by ROSAT, the crosses indicate $L_{X}/L_{bol}$ upper limits for the component of binaries with no X-ray emission detected. The dotted line shows the best least square fit to the data (open circles, M $\leq$ 0.77$M_{\sun}$). The dot-dashed line shows the best least square fit to the data including stars with M $>$ 0.77$M_{\sun}$. The dashed line in both figures indicates the model's zero-deviation baseline.}
\label{fig:1Ms_01_01}
\end{figure} 

\begin{figure}[t]
\epsscale{1.0}
\plotone{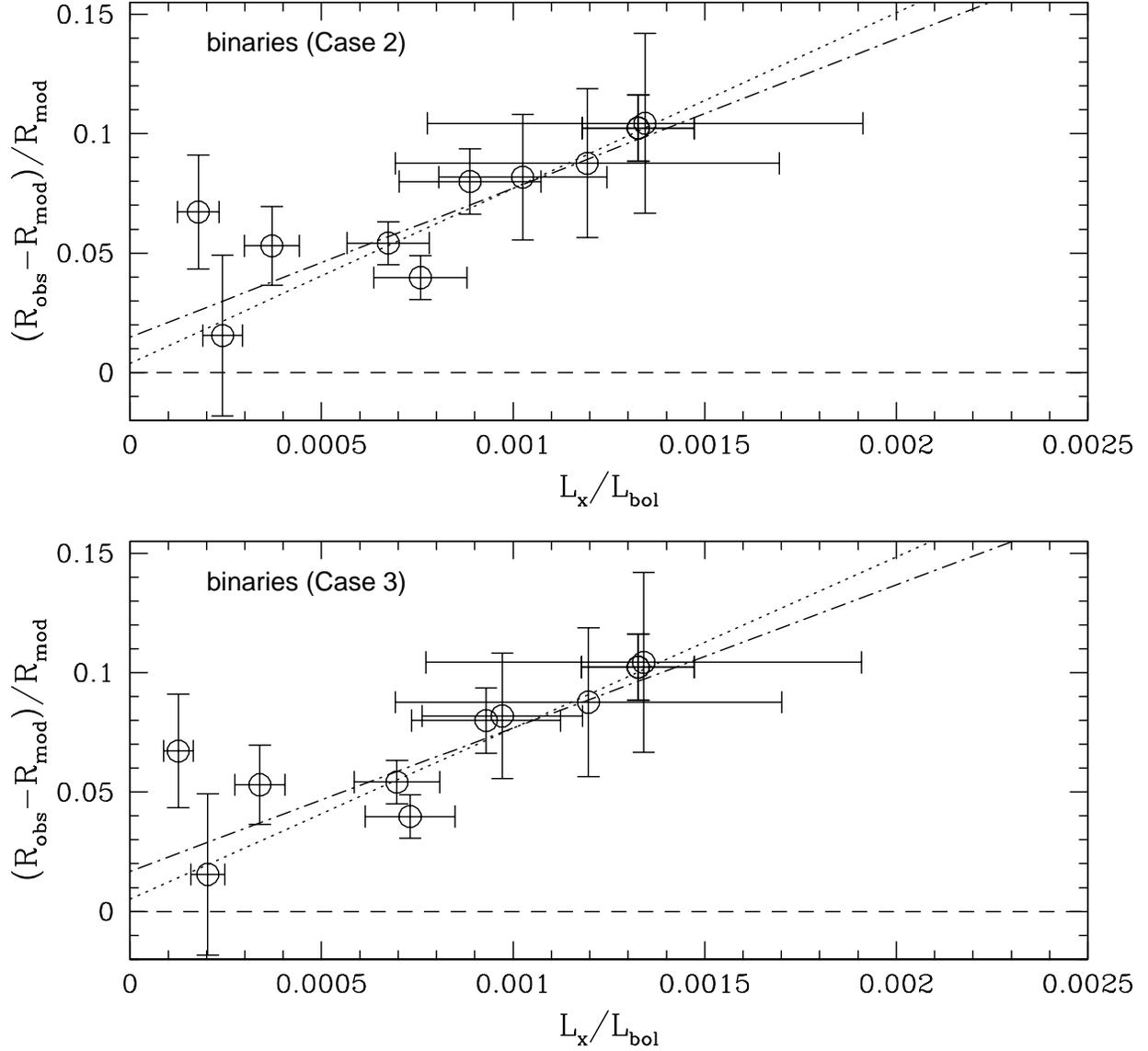}
\caption{${\it Top}$ -- Fractional deviation of the radii of the binary stars in the sample with detected X-ray emission from the 1Gyr, Z = 0.02, $\alpha$ = 1.0 Baraffe et al. (1998) model as a function of $L_{X}/L_{bol}$ for {\it Case 2}, i.e. assuming $L_{X}/L_{bol}$ $\propto$ $v_{rot}$sini. ${\it Bottom}$ -- Same as top figure for {\it Case 3}, i.e. assuming $L_{X}/L_{bol}$ $\propto$ $(v_{rot}sini)^{2}$. The dotted lines show the best least square fits to the data with M $\leq$ 0.77$M_{\sun}$. The dot-dashed lines show the best least square fit to the data including stars with M $>$ 0.77$M_{\sun}$. The dashed lines show the model's zero-deviation baseline.}
\label{fig:1Ms_01_02}
\end{figure} 

\begin{figure}[t]
\epsscale{1.0}
\plotone{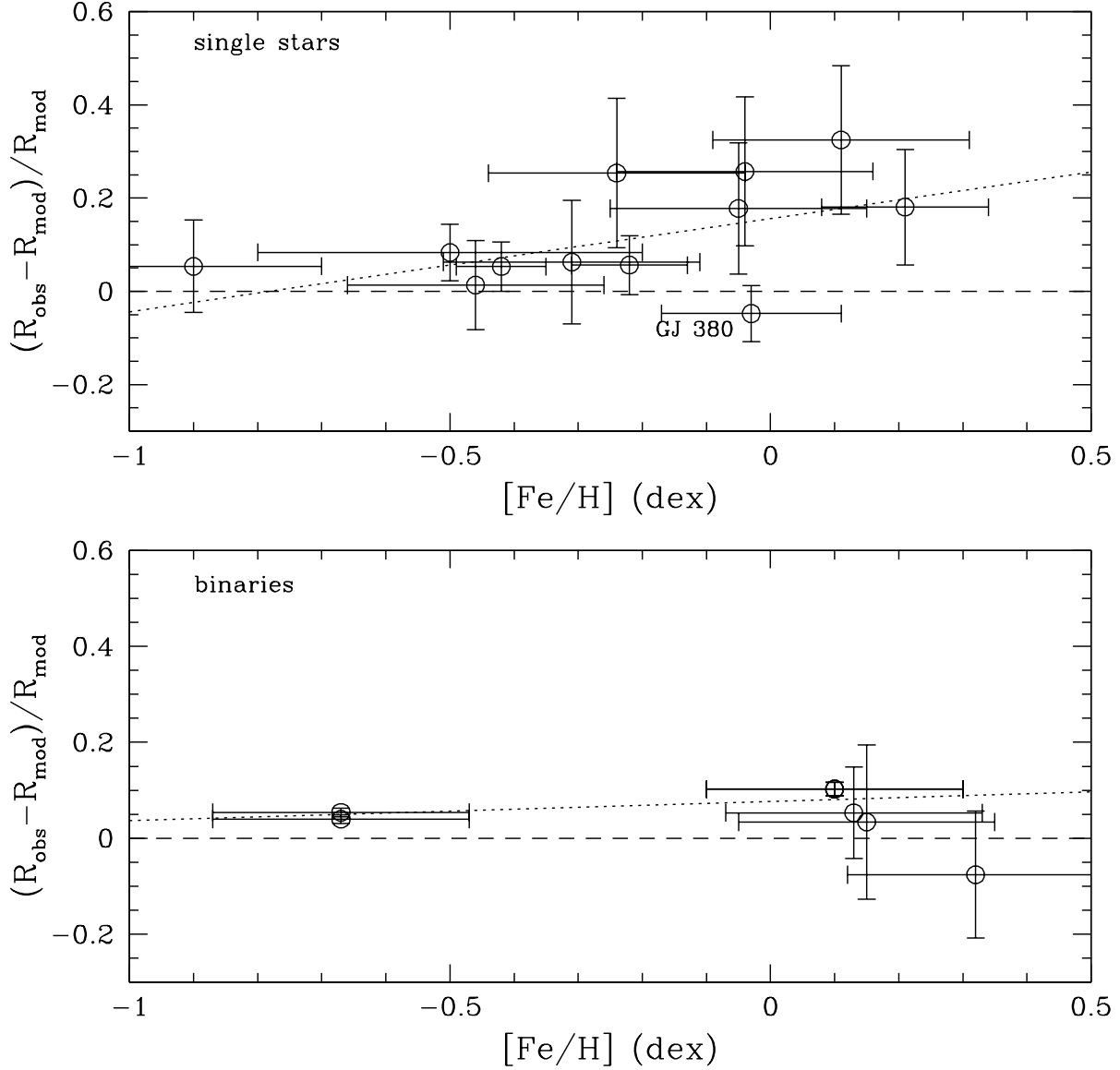}
\caption{Fractional deviation of the radius of stars in the sample from the 1Gyr, Z = 0.02, $\alpha$ = 1.0 Baraffe et al. (1998) model as a function of metallicity. The {\it Top} diagram shows the single stars. The {\it Bottom} diagram shows the binary low-mass secondaries V818 Tau B, OGLE-TR-34 B, and OGLE-TR-122 B, and the M-dwarf binaries YY Gem and CM Dra. The dotted lines show the best least square fits to the data. The dashed lines show the model's zero-deviation baseline.}
\label{fig:1Ms_02}
\end{figure} 

\end{document}